\documentclass[epj,nopacs]{svjour}
\usepackage{graphics}
\begin{document}
\title{Higher harmonics of azimuthal anisotropy in 
relativistic heavy ion collisions in HYDJET++ model}
\author{L.V.~Bravina\inst{2}, B.H.~Brusheim Johansson\inst{2,3}, G.Kh.~Eyyubova\inst{1,4}, 
V.L.~Korotkikh\inst{1}, I.P.~Lokhtin\inst{1}, L.V.~Malinina\inst{1}, 
S.V.~Petrushanko\inst{1}, A.M.~Snigirev\inst{1}, E.E.~Zabrodin\inst{2,1}}  
%
%
\institute {Skobeltsyn Institute of Nuclear Physics, Lomonosov Moscow State 
University, Moscow, Russia \and 
The Department of Physics, University of Oslo, Norway \and
Oslo and Akershus University College for Applied Sciences, Oslo, Norway \and 
Czech Technical University in Prague, FNSPE, Prague, Czech Republic}
%
%

\abstract{
The LHC data on azimuthal anisotropy harmonics from PbPb collisions 
at center-of-mass energy 2.76 TeV per nucleon pair are analyzed and 
interpreted in the framework of the HYDJET++ model. 
The cross-talk of elliptic $v_2$ and triangular $v_3$ flow in
the model generates both even and odd harmonics of higher order. 
Comparison with the experimental data shows that this mechanism is
able to reproduce the $p_{\rm T}$ and centrality dependencies of
quadrangular flow $v_4$, and also the basic trends for 
pentagonal $v_5$ and hexagonal $v_6$ flows.   
%
}
%
\titlerunning{Higher harmonics of azimuthal anisotropy...}
\authorrunning{L.V. Bravina et al.}
\maketitle

\section{Introduction}
\label{sec:intro}

The study of the fundamental theory of strong interactions (Quantum 
Chromodynamics, QCD) in the regimes of extreme densities and temperatures 
is ongoing via the measurement of the properties of hot and dense 
multi-parton systems produced in high-energy nuclear collisions (see, e.g., 
reviews~\cite{d'Enterria:2006su,Hwa:2010,Salgado:2009jp,Dremin:2010jx}).
The started LHC heavy ion program makes it possible to probe the new 
frontiers of the high temperature QCD providing the valuable information 
on the dynamical behavior of quark-gluon matter (QGM), as predicted by 
lattice calculations. A number of interesting LHC results from 
PbPb runs at $\sqrt s_{\rm NN}=2.76$ TeV, have been published by ALICE, 
ATLAS and CMS collaborations (see~\cite{Muller:2012zq} for the overview 
of the results from the first year of heavy ion physics at LHC). 

One of the modern trends in heavy ion physics at high energies is a study of Fourier harmonics 
of azimuthal particle distribution, which is a powerful probe of bulk properties of 
the created high density matter. It is typically described by a Fourier series of the form: 
\begin{eqnarray}
\displaystyle
\label{eq:1}
& & E\frac{d^3N}{dp^3}=\frac{d^2N}{2\pi p_{\rm T}dp_{\rm T}d\eta} \times 
\nonumber \\
& & \{
1+2\sum\limits_{n = 1}^\infty v_{\rm n}(p_{\rm T},\eta)  
\cos{ \left[ n(\varphi -\Psi_{\rm n}) \right] }
\} ~,
\end{eqnarray}
where $\varphi$ is the azimuthal angle with respect to the reaction plane $\Psi_{\rm n}$, 
and $v_{\rm n}$ are the Fourier coefficients. The second harmonic, $v_2$, referred to as 
``elliptic flow'', is the most extensively studied one, because it directly relates the 
anisotropic shape of the overlap of the colliding nuclei to the corresponding anisotropy of 
the outgoing momentum distribution. The momentum and centrality dependencies of the elliptic 
flow in PbPb collisions were measured at the 
LHC~\cite{Aamodt:2010pa,ATLAS:2011yk,Chatrchyan:2012ta} in the first instance. Then, the 
results of measurements of the higher azimuthal 
harmonics~\cite{ALICE:2011ab,Aad:2012bu,Chatrchyan:2013kba} and the anisotropic flow of 
identified particles~\cite{Abelev2012:di} were published. The higher order coefficients 
$v_{\rm n}$ (n$>2$) are smaller than $v_2$. They also carry important information on the 
dynamics of the medium created, and complement $v_2$ in providing a more complete picture of 
its bulk properties. The two coefficients that have been closely studied are the quadrangular 
(or hexadecapole) flow $v_4$~\cite{Kolb:2003zi,Kolb:2004gi} and triangular flow 
$v_3$~\cite{Alver:2010gr}. Although the pentagonal and hexagonal flows $v_5$ and $v_6$ are 
studied to a lesser extent, there exist some predictions from hydrodynamics on them 
also~\cite{Alver:2010dn}. 

At relatively low transverse momenta, $p_{\rm T}<3\div4$ GeV/$c$, the azimuthal anisotropy 
results from a pressure-driven anisotropic expansion of the created matter, with more particles 
emitted in the direction of the largest pressure gradients~\cite{Ollitrault:1992bk}. At higher 
$p_{\rm T}$, this anisotropy is understood to result from the path-length dependent energy loss 
of partonic jets as they traverse the matter, with more jet particles emitted in the direction 
of shortest path-length~\cite{Gyulassy:2000gk}.

In Ref.~\cite{Lokhtin:2012re} the LHC data on multiplicity, charged hadron spectra, elliptic 
flow and femtoscopic correlations from PbPb collisions were analyzed in the frameworks of the 
HYDJET++ model~\cite{Lokhtin:2008xi}. Taking into account both hard and soft components and 
tuning input parameters allow HYDJET++ to reproduce these data. Another 
study \cite{Bravina:2013upa} with HYDJET++ was dedicated to the influence of jet production 
mechanism on the ratio $v_4/v_2^2$ and its role in violation of the number-of-constituent-quark 
(NCQ) scaling~\cite{Noferini:2012ps}, predicted within the HYDJET++ in~\cite{Eyyubova:2009hh}.
In the current paper, tuned HYDJET++ is applied to analyze the LHC data on momentum and 
centrality dependences of azimuthal anisotropy harmonics in PbPb collisions, and then to 
illuminate the mechanisms of the generation of Fourier coefficients $v_2 \div v_6$. The 
detailed study of hexagonal flow $v_6$ is also the subject of our recent 
paper~\cite{Bravina:2013ora}. 

Note that the LHC data on higher-order azimuthal aniso\-tropy harmonics ($v_2 \div v_4$) were 
analyzed with a multiphase transport model (AMPT) in~\cite{Xu:2011jm}. It was shown that AMPT 
describes LHC data on the anisotropic flow coefficients $v_{\rm n}$ (n=2$\div$4) for 
semi-central PbPb collisions at $p_{\rm T} < 3$ GeV/$c$. It also reproduces reasonably well the 
centrality dependence of integral $v_{\rm n}$ for all but most central collisions. Another 
approach~\cite{Gale:2012rq} reproducing $v_{\rm n}$ data in ultrarelativistic heavy ion 
collisions is the glasma flow with the subsequent relativistic viscous hydrodynamic evolution of 
matter through the quark-gluon plasma and hadron gas phases (IP-Glasma+MUSIC model). This model 
gives good agreement to $p_{\rm T}$-dependence of $v_{\rm n}$ (n=2$\div$5) and event-by-event 
distributions of $v_2 \div v_4$ at RHIC and LHC.

The study of generation of higher flow harmonics within the 
HYDJET++ has several attractive features. Firstly, the presence of
elliptic and triangular flow permits us to examine the interference 
of these harmonics and its contribution to all higher even and odd 
components of the anisotropic flow. If necessary, the original 
eccentricities of higher order can be easily incorporated in the
model for the fine tuning of the distributions. Secondly, very rich
table of resonances, which includes about 360 meson and baryon species,
helps one to analyze all possible final state interactions. Thirdly,
the interplay of ideal hydrodynamics with jets can unveil the role of
hard processes in the formation of anisotropic flow of secondary 
hadrons. The basic features of the model are described in Sect.~\ref{sec:model}. 

\section{HYDJET++ model}
\label{sec:model}

HYDJET++ (the successor of HYDJET~\cite{Lokhtin:2005px}) is the
model of relativistic heavy ion collisions, which includes two
independent components: the soft state (hydro-type) and the hard
state resulting from the in-medium multi-parton fragmentation. The
details of the used physics model and simulation procedure can be
found in the HYDJET++ manual~\cite{Lokhtin:2008xi}. Main features 
of the model are sketched below as follows.

The soft component of an event in HYDJET++ is the ``thermal''
hadronic state generated on the chemical and thermal freeze-out
hypersurfaces obtained from the pa\-ra\-met\-ri\-za\-ti\-on of
relativistic hydrodynamics with preset freeze-out conditions (the
adapted event generator FAST MC~\cite{Amelin:2006qe,Amelin:2007ic}). 
Hadron multiplicities are calculated using the effective thermal volume 
approximation and Poisson multiplicity distribution around its mean value, 
which is supposed to be proportional to a number of participating nucleons
for a given impact parameter of a AA collision. To simulate the
elliptic flow effect, the hydro-inspired pa\-ra\-met\-ri\-za\-ti\-on is
implemented for the momentum and spatial anisotropy of a soft
hadron emission source~\cite{Lokhtin:2008xi,Wiedemann:1997cr}.

The model used for the hard component in HYDJET++ is based on the 
PYQUEN partonic energy loss model~\cite{Lokhtin:2005px}. The approach 
describing the multiple scattering of hard partons relies on accumulated
energy loss via gluon radiation which is associated with each
parton scattering in expanding quark-gluon fluid. It also includes
the interference effect in gluon emission with a finite formation
time using the modified radiation spectrum $dE/dx$ as a function
of the decreasing temperature $T$. The model takes into account
radiative and collisional energy loss of hard partons in
longitudinally expanding quark-gluon fluid, as well as the
realistic nuclear geometry. The simulation of single hard nucleon-nucleon 
sub-collisions by PYQUEN is constructed as a modification of the jet event 
obtained with the generator of hadron-hadron interactions
PYTHIA$\_$6.4~\cite{Sjostrand:2006za}. Note, that Pro-Q20 tune was used for 
the present simulation. The number of PYQUEN jets is generated according 
to the binomial distribution. The mean number of jets produced in an AA
event is calculated as a product of the number of binary NN
sub-collisions at a given impact parameter per the integral cross
section of the hard process in NN collisions with the minimum
transverse momentum transfer $p_{\rm T}^{\rm min}$ (the latter is an
input parameter of the model). In HYDJET++, partons
produced in (semi)hard processes with the momentum transfer lower
than $p_{\rm T}^{\rm min}$, are considered as being ``thermalized''. 
So, their hadronization products are included ``automatically'' in the
soft component of the event. In order to take into account the
effect of nuclear shadowing on parton distribution functions, we
use the impact parameter dependent 
pa\-ra\-met\-ri\-za\-ti\-on~\cite{Tywoniuk:2007xy} obtained in the framework 
of Glauber-Gribov theory. 

The model has a number of input parameters for the soft and hard components.
They are tuned from fitting to experimental data values for various
physical observables, see~\cite{Lokhtin:2012re} for details. 

In order to simulate higher azimuthal anisotropy harmonics, the following modification has been 
implemented in the model. HYDJET++ does not contain the fireball evolution from the initial 
state to the freeze-out stage. Instead of application of computational relativistic 
hydrodynamics, which is extremely time consuming, HYDJET++ employs the simple and frequently 
used parametrizations of the freeze-out hypersurface~\cite{Lokhtin:2008xi}. Then, anisotropic 
elliptic shape of the initial overlap of the colliding nuclei results in a corresponding 
anisotropy of the outgoing momentum distribution. To describe the second harmonic $v_2$ the 
model utilizes coefficients $\delta(b)$ and $\epsilon(b)$ representing, respectively, the flow 
and the coordinate anisotropy of the fireball at the freeze-out stage as functions of the 
impact parameter $b$. These momentum and spatial anisotropy parameters $\delta(b)$ and 
$\epsilon(b)$ can either be treated independently for each centrality, or can be related to 
each other through the dependence on the initial ellipticity $\epsilon_0(b)=b/2R_A$, where 
$R_A$ is the nucleus radius. The last option allows us to describe the elliptic flow 
coefficient $v_2$ for most centralities at the RHIC~\cite{Lokhtin:2008xi} and 
LHC~\cite{Lokhtin:2012re} energies using only two independent on centrality parameters.
 
\begin{figure*}
\begin{center}
\resizebox{1.\textwidth}{!}{%
\includegraphics{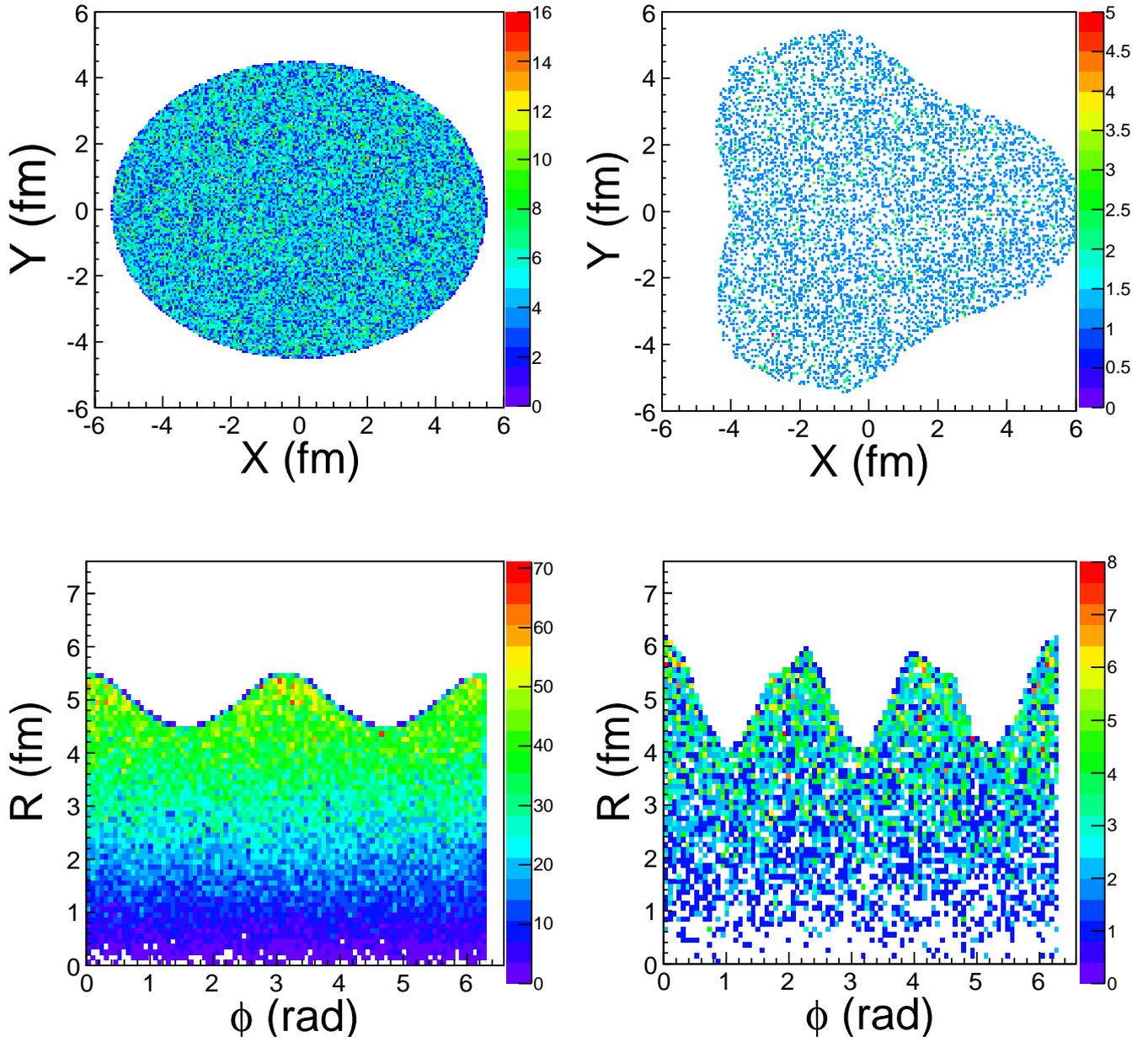}
}
\end{center}
 \caption{The carton figure illustrating the second (left, $\epsilon_{3}(b)=0$) and 
second+third (right, $\epsilon_{3}(b)=0.2$) azimuthal anisotropy harmonics generation in 
HYDJET++ at $R_{\rm f}(b)=5$ fm, $\epsilon(b)=-0.2$, $\Psi_2^{\rm RP}=0$, $\Psi_3^{\rm RP}=0$. 
Particle densities in the transverse plane are shown for X-Y (top) and R-$\phi$ (bottom) 
representations.} 
\label{XY_HY}
\vskip 2 cm
\end{figure*}

Non-elliptic shape of the initial overlap of the colliding nuclei, which can be characterized 
by the initial triangular coefficient $\epsilon_{03}(b)$, results in an appearance of higher 
Fourier harmonics in the outgoing momentum distribution. Our Monte-Carlo (MC) procedure allows 
us to parametrize easily this anisotropy via the natural modulation of final freeze-out 
hypersurface, namely
\begin{equation}
\label{Rbphi}
R(b,\phi)= R_{\rm f}(b)
\frac{\sqrt{1-\epsilon^2(b)}}{\sqrt{1+\epsilon(b)\cos2\phi}}[1+\epsilon_3(b)
\cos3(\phi+\Psi_3^{\rm RP})]~,
\end{equation} 
where $\phi$ is the spatial azimuthal angle of the fluid element relatively to 
the direction of the impact parameter. $R(b,\phi)$ is the fireball transverse radius in 
the given azimuthal direction $\phi$ with the scale $R_{\rm f}(b)$, which is a model 
parameter. The phase $\Psi_3^{\rm RP}$ allows us to introduce the third harmonics 
possessing its own reaction plane, randomly distributed with respect to the direction of 
the impact parameter ($\Psi_2^{\rm RP}=0$). This new anisotropy parameter $\epsilon_3(b)$ can 
again be treated independently for each centrality, or can be expressed through the initial 
ellipticity $\epsilon_0(b)=b/2R_A$. Note, that such modulation does not affect the elliptic 
flow coefficient $v_2$, which was fitted earlier with two parameters $\delta(b)$ and 
$\epsilon(b)$~\cite{Lokhtin:2012re,Lokhtin:2008xi}. Figure~\ref{XY_HY} illustrates second and 
third harmonics generation in HYDJET++ by representing particle densities in the transverse 
plane. One should be aware that the triangular deformation shown here is very strong. The 
actual deformations needed to describe triangular flow at LHC energies are typically order of 
magnitude weaker.

The modulation of the maximal transverse flow rapidity, first considered in Eq.~(28)
of Ref.~\cite{Lokhtin:2008xi} at the paramet\-ri\-za\-tion of 4-velocity $u$,
\begin{eqnarray}
\label{v34}
\rho_{\rm u}^{\rm max}= \rho_{\rm u}^{\rm max}(b=0)[1+ \rho_{\rm 3u}(b) \cos3\phi +
\rho_{\rm 4u}(b) \cos4\phi]~, 
\end{eqnarray}
also permits the introduction of higher azimuthal harmonics related, however, to the direction
of the impact parameter ($\Psi_2^{\rm RP}=0$) only. In this case we get the modulation of the 
velocity profile in all freeze-out hypersurface, and can not ``rotate'' this modulation with 
independent phase. The new anisotropy parameters, $\rho_{\rm 3u}(b)$ and  $\rho_{\rm 4u}(b)$, 
can again be treated independently for each centrality, or can be expressed through the initial 
ellipticity $\epsilon_0(b)=b/2R_A$. 

For current simulations we have introduced the minimal modulation in HYDJET++ using just simple  
parameterizations $\epsilon_3(b)\propto \epsilon_0^{1/3}(b)$ and 
$\rho_{\rm 4u}(b)\propto \epsilon_0(b)$, while $\rho_{\rm 3u}(b)$ being taken equal to zero.
The corresponding proportionality factors were selected from the best fit of the data to 
$v_3(p_{\rm T})$ and $v_4(p_{\rm T})$. 

Let us mark that the azimuthal anisotropy parameters $\epsilon(b)$,  
$\delta(b)$ and $\epsilon_3(b)$ are fixed at given impact parameter b. 
Therefore they do not provide dynamical event-by-event flow fluctuations, 
and specify $v_{\rm n}(b)$ accumulated over many events. The main source 
of flow fluctuations in HYDJET++ is fluctuations of particle momenta and 
multiplicity. Recall, that the momentum-coordinate correlations in 
HYDJET++ for soft component is governed by collective velocities of fluid
elements, and so the fluctuations in particle coordinates are reflected in 
their momenta. The fluctuations became stronger as resonance decays and
(mini-)jet production are taken into account. An event distribution over
collision impact parameter for each centrality class also increases such 
fluctuations. In the current paper we restrict ourselves to analysis of the 
event-averaged $v_{\rm n}(p_{\rm T})$. The detailed study of event-by-event 
flow fluctuations is the subject of our future investigation. The possible
further modification of HYDJET++ to match experimental data on flow 
fluctuations would be smearing of parameters $\epsilon$, $\delta$ and 
$\epsilon_3$ at a given b.
 
\section{Results}
\label{sec:results}

It was demonstrated in~\cite{Lokhtin:2012re} that tuned HYDJET++ model can reproduce the LHC 
data on centrality and pseudorapidity dependence of inclusive charged particle multiplicity, 
$p_{\rm T}$-spectra and $\pi^\pm \pi^\pm$ correlation radii in central PbPb collisions, and 
$p_{\rm T}$- and $\eta$-dependencies of the elliptic flow coefficient $v_2$  (up to $p_{\rm T} 
\sim 5$ GeV/$c$ and 40\% centrality). However the reasonable treatment of higher and odd Fourier 
harmonics of particle azimuthal distribution $v_{\rm n}$ ($n>2$) needs the additional modifications of 
the model, which does not effect azimuthally-integrated physical observables (see previous 
section). We have compared the results of HYDJET++ simulations with the LHC data on 
$v_{\rm n}$ for inclusive as well as for identified charged hadrons.

\subsection{Anisotropy harmonics for inclusive charge hadrons}
\label{subsec:res1}

The standard way of measuring $v_{\rm n}$ corresponds to the inclusive 
particle harmonics on the base of Eq.~(\ref{eq:1}). 
Then $v_{\rm n}$ is extracted using the special methods, such as the event plane 
$v_{\rm n}\{\rm EP\}$~\cite{Poskanzer:1998yz}, or $m$-particle cumulant 
$v_{\rm n}\{m\}$~\cite{Borghini:2001vi,Borghini:2001zr}, or Lee-Yang zero 
methods $v_{\rm n}\{\rm LYZ\}$~\cite{Bhalerao:2003xf,Borghini:2004ke}. 
In order to estimate the uncertainties related to the experimental definitions of flow 
harmonics, HYDJET++ results for different methods of $v_{\rm n}$ extraction were compared with 
its ``true'' values, known from the event generator and determined relatively to 
$\Psi_2^{\rm RP}$ for even and $\Psi_3^{\rm RP}$ for odd harmonics, respectively. 

\begin{figure*}
\begin{center}
\resizebox{0.85\textwidth}{!}{%
\includegraphics{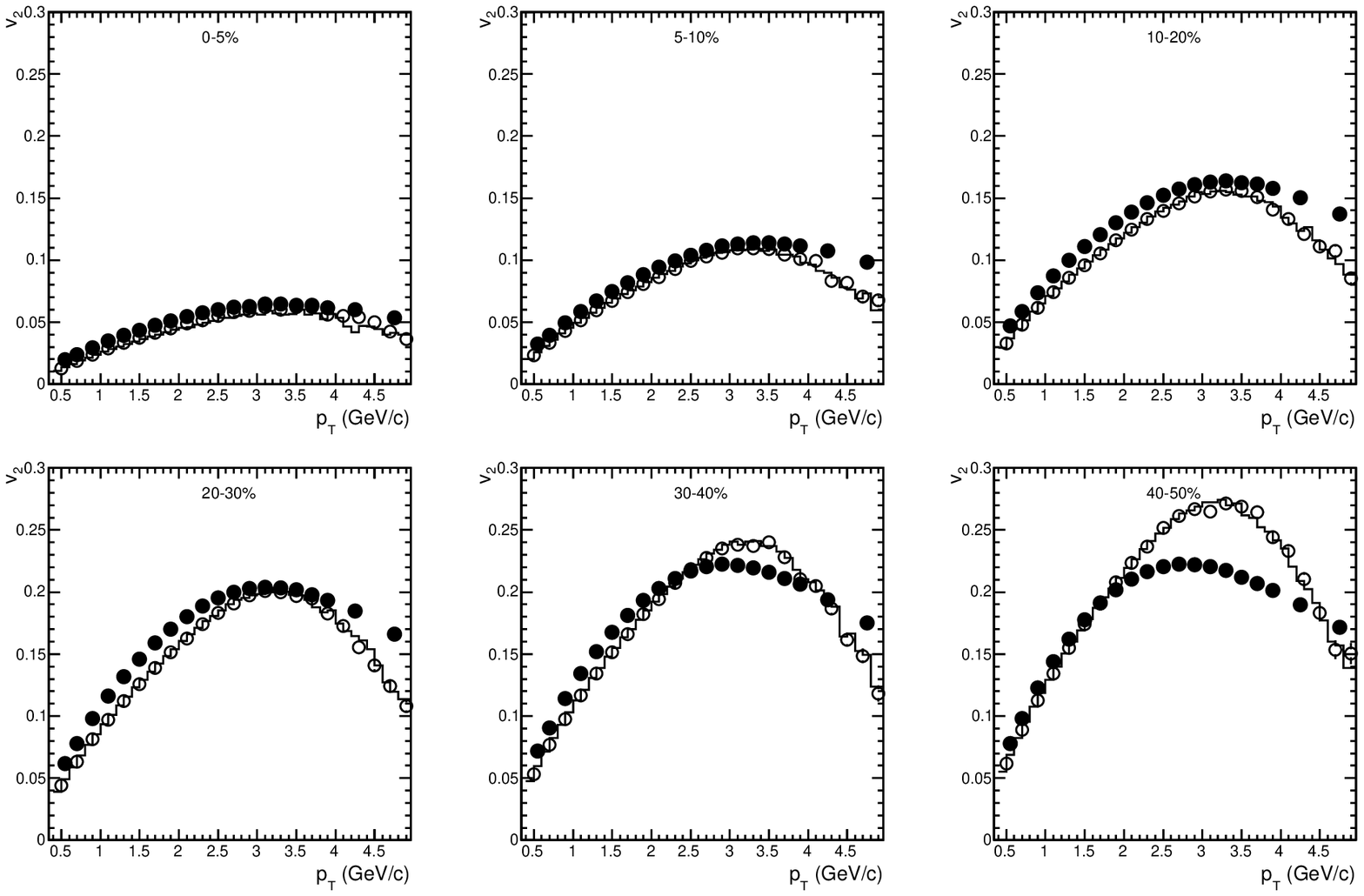}
}
\end{center}
 \caption{Elliptic flow $v_2(p_{\rm T})$ of charged hadrons at pseudo-rapidity $|\eta|<2.5$ for 
different centralities of PbPb collisions at $\sqrt s_{\rm NN}=2.76$ TeV. The closed circles 
are ATLAS data~\cite{Aad:2012bu} on $v_2\{\rm EP\}$, open circles and histograms represent  
$v_2\{\rm EP\}$ and $v_2(\Psi_2^{\rm RP})$ for HYDJET++ events, respectively.} 
\label{v2_ATLAS}
\end{figure*}

\begin{figure*}
\begin{center}
\resizebox{0.85\textwidth}{!}{%
\includegraphics{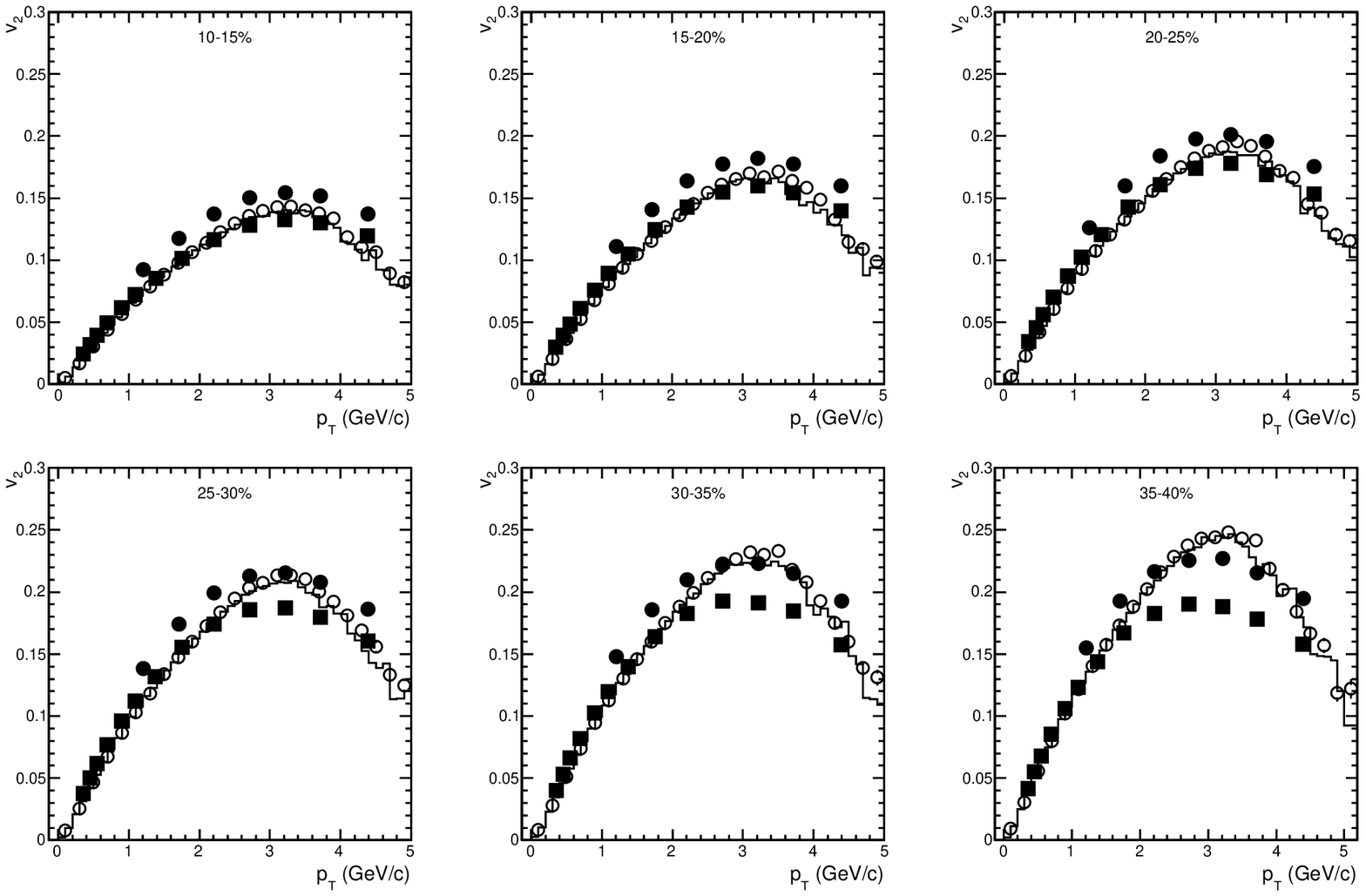}
}
\end{center}
 \caption{Elliptic flow $v_2(p_{\rm T})$ of charged hadrons at pseudo-rapidity $|\eta|<0.8$ for 
different centralities of PbPb collisions at $\sqrt s_{\rm NN}=2.76$ TeV. The closed points are 
CMS data~\cite{Chatrchyan:2012ta} ($v_2\{2\}$ --- circles, $v_2\{\rm LYZ\}$ --- squares), 
open circles and histograms represent $v_2\{\rm EP\}$ and $v_2(\Psi_2^{\rm RP})$ for HYDJET++ 
events, respectively.} 
\label{v2_CMS}
\end{figure*}

\begin{figure*}
\begin{center}
\resizebox{0.85\textwidth}{!}{%
\includegraphics{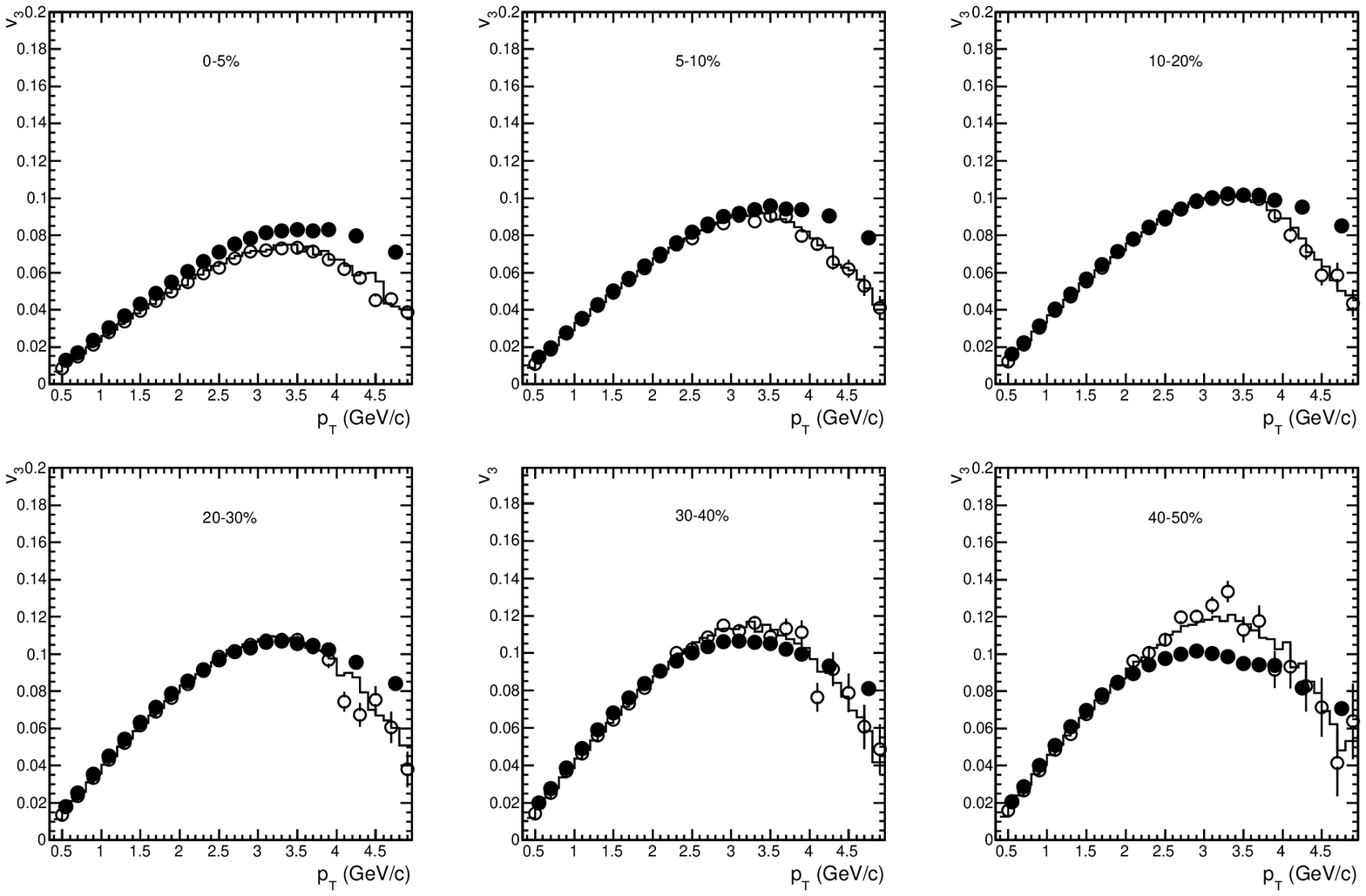}
}
\end{center}
 \caption{Triangular flow $v_3(p_{\rm T})$ of charged hadrons at pseudo-rapidity $|\eta|<2.5$ 
for different centralities of PbPb collisions at $\sqrt s_{\rm NN}=2.76$ TeV. The closed 
circles are ATLAS data~\cite{Aad:2012bu} on $v_3\{\rm EP\}$, open circles and histograms 
represent $v_3\{\rm EP\}$ and $v_3(\Psi_3^{\rm RP})$ for HYDJET++ events, respectively.}  
\label{v3_ATLAS}
\end{figure*}

\begin{figure*}
\begin{center}
\resizebox{0.85\textwidth}{!}{%
\includegraphics{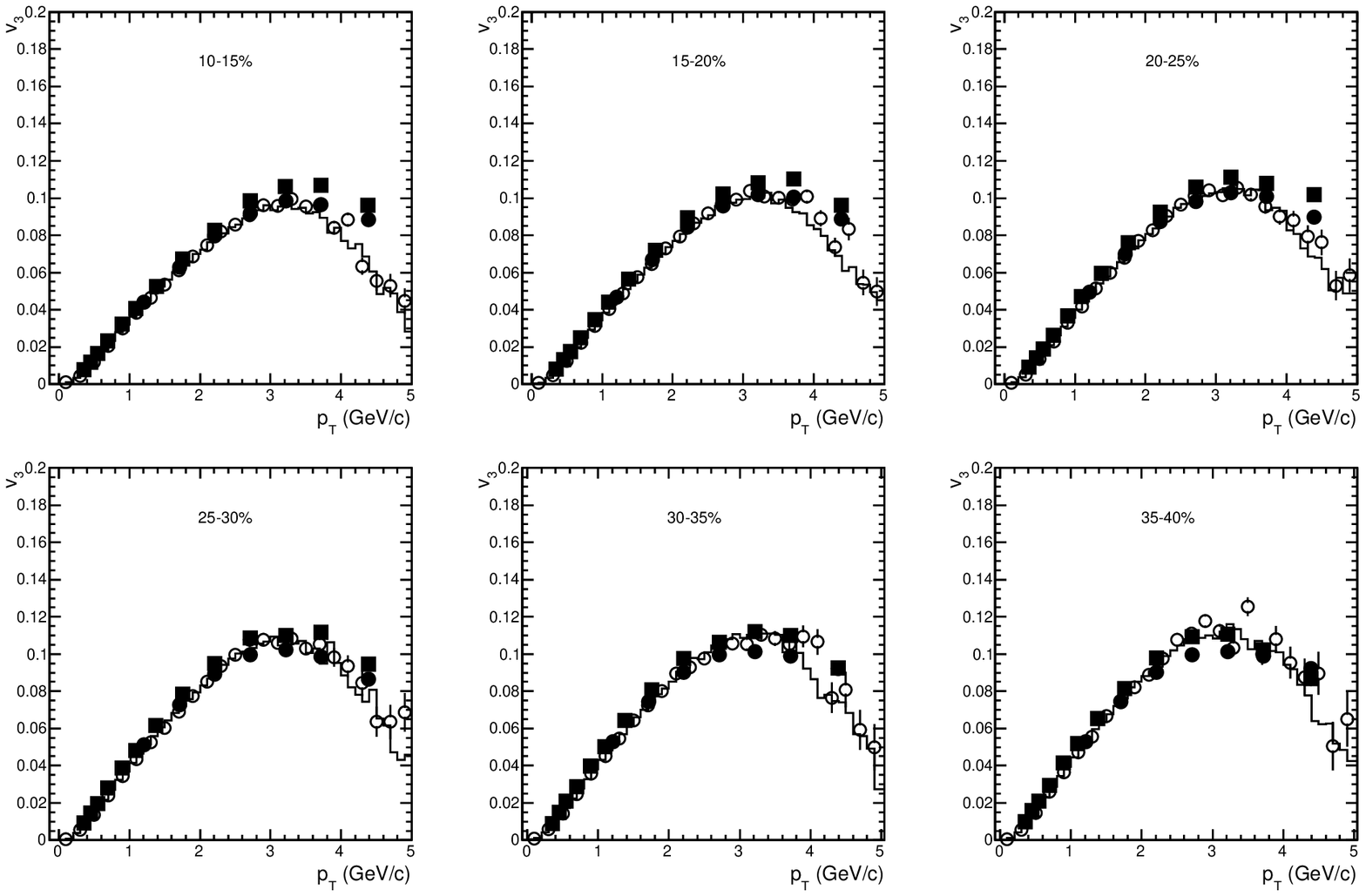}
}
\end{center}
 \caption{Triangular flow $v_3(p_{\rm T})$ of charged hadrons at pseudo-rapidity $|\eta|<0.8$ 
for different centralities of PbPb collisions at $\sqrt s_{\rm NN}=2.76$ TeV. The closed points 
are CMS data~\cite{Chatrchyan:2013kba} ($v_3\{2\}$ --- circles, $v_3\{\rm EP\}$ --- squares), 
open circles and histograms represent $v_3\{\rm EP\}$ and $v_3(\Psi_3^{\rm RP})$ for HYDJET++ 
events, respectively.} 
\label{v3_CMS}
\end{figure*}

\begin{figure*}
\begin{center}
\resizebox{0.85\textwidth}{!}{%
\includegraphics{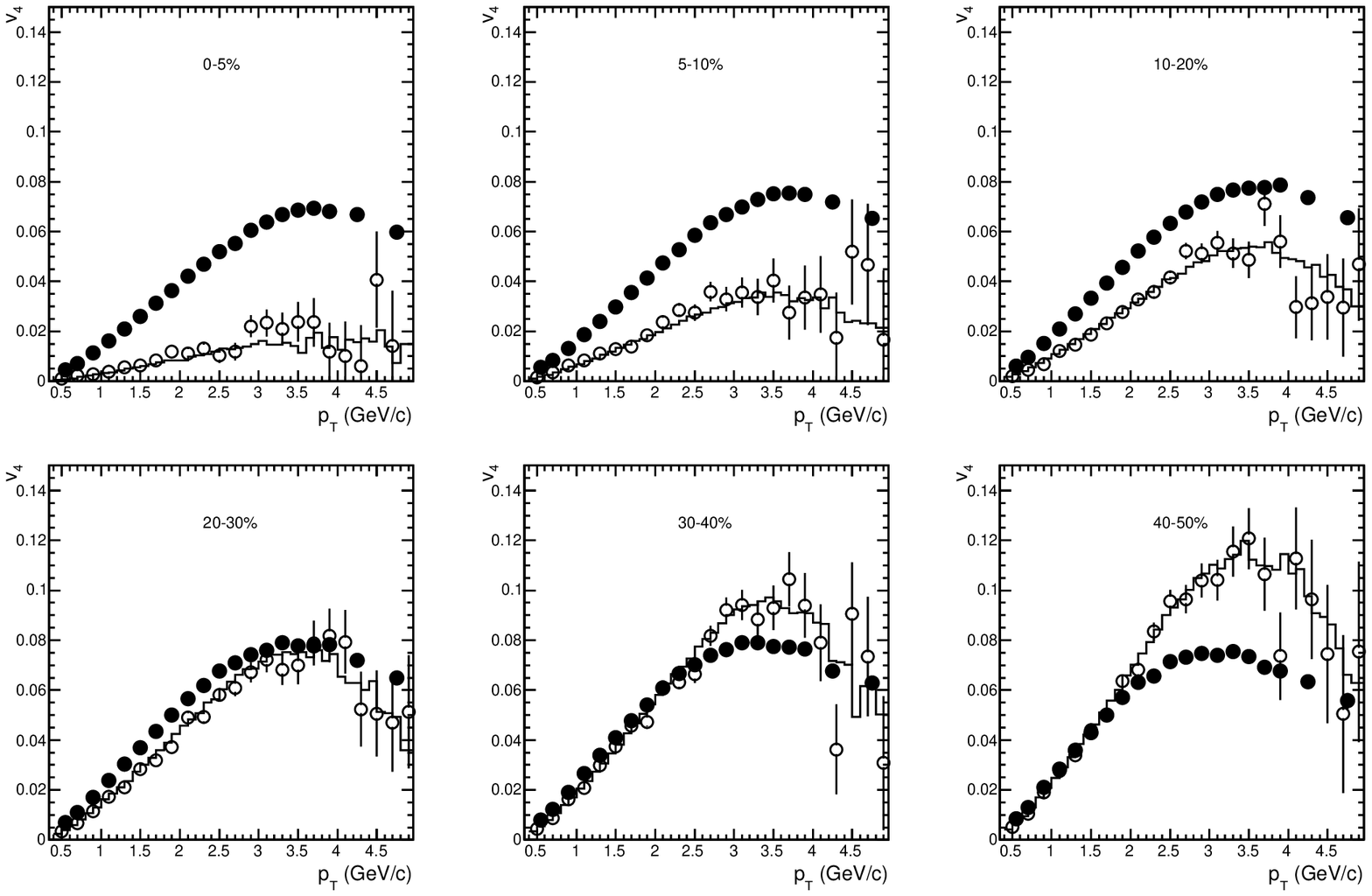}
}
\end{center}
  \caption{Quadrangular flow $v_4(p_{\rm T})$ of charged hadrons at pseudo-rapidity $|\eta|<2.5$ 
for different centralities of PbPb collisions at $\sqrt s_{\rm NN}=2.76$ TeV. The closed circles 
are ATLAS data~\cite{Aad:2012bu} on $v_4\{\rm EP\}$, open circles and histograms represent  
$v_4\{\rm EP\}$ and $v_4(\Psi_2^{\rm RP})$ for HYDJET++ events, respectively.} 
\label{v4_ATLAS}
\end{figure*}

\begin{figure*}
\begin{center}
\resizebox{0.85\textwidth}{!}{%
\includegraphics{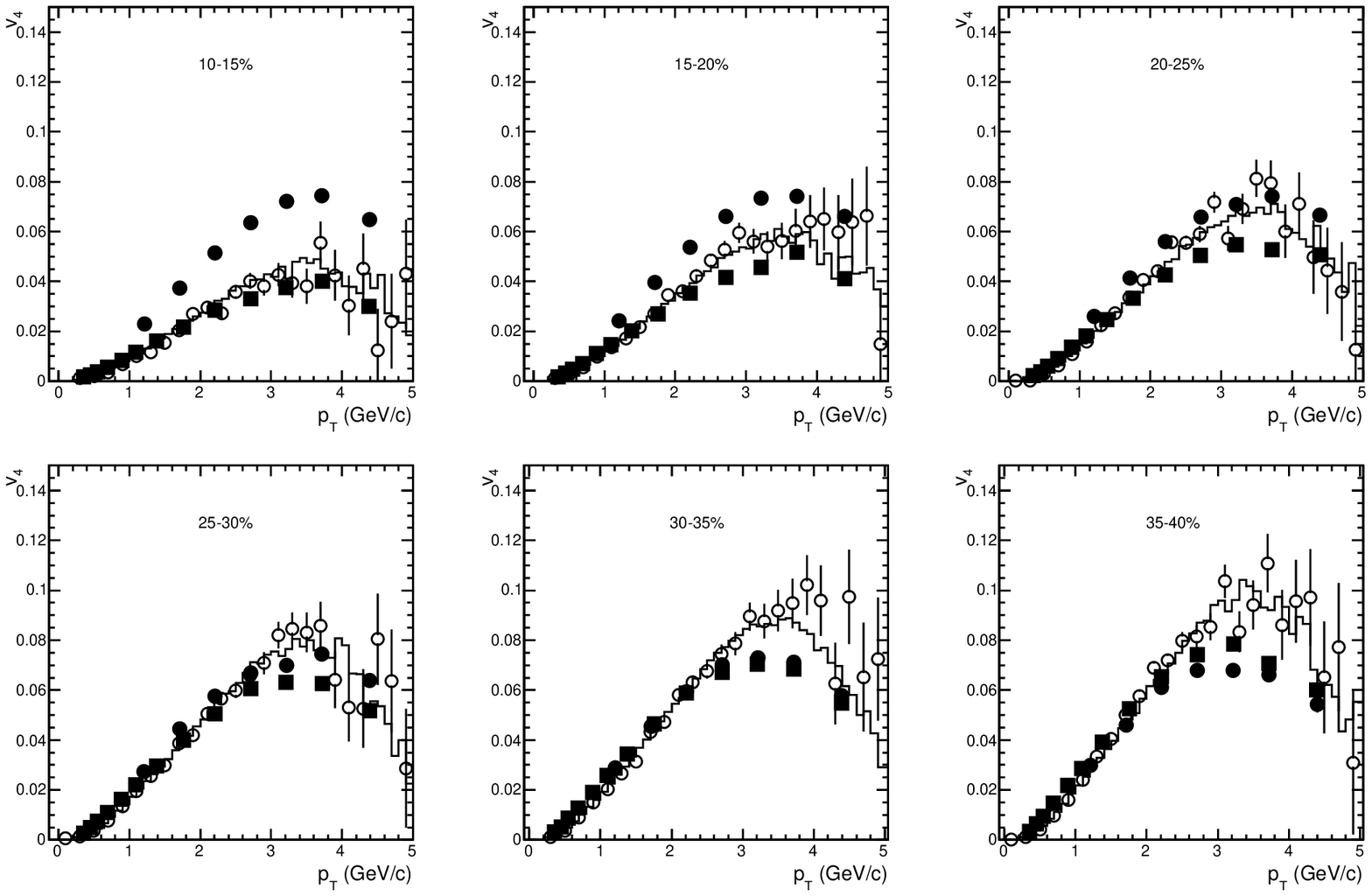}
}
\end{center}
 \caption{Quadrangular flow $v_4(p_{\rm T})$ of charged hadrons at pseudo-rapidity $|\eta|<0.8$ 
for different centralities of PbPb collisions at $\sqrt s_{\rm NN}=2.76$ TeV. The closed points 
are CMS data~\cite{Chatrchyan:2013kba} ($v_4\{2\}$ --- circles, $v_4\{\rm LYZ\}$ --- squares), 
open circles and histograms represent $v_4\{\rm EP\}$ and $v_4(\Psi_2^{\rm RP})$ for HYDJET++ 
events, respectively.} 
\label{v4_CMS}
\end{figure*}

\begin{figure*}
\begin{center}
\resizebox{0.85\textwidth}{!}{%
\includegraphics{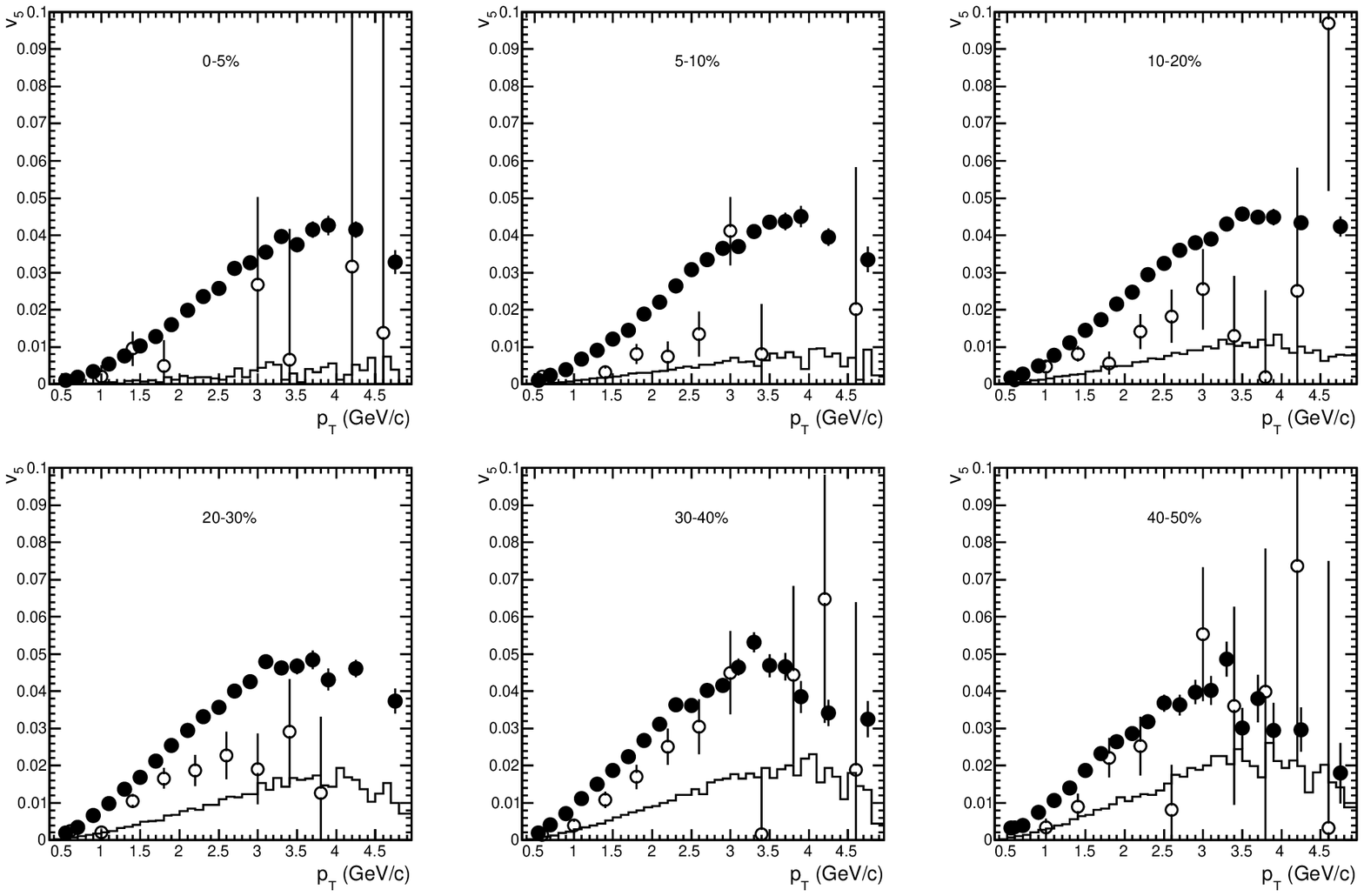}
}
\end{center}
 \caption{Pentagonal flow $v_5(p_{\rm T})$ of charged hadrons at pseudo-rapidity $|\eta|<2.5$ for 
different centralities of PbPb collisions at $\sqrt s_{\rm NN}=2.76$ TeV. The closed circles 
are ATLAS data~\cite{Aad:2012bu} on $v_5\{\rm EP\}$, open circles and histograms represent 
$v_5\{\rm EP\}$ and $v_5(\Psi_3^{\rm RP})$ for HYDJET++ events, respectively.} 
\label{v5_ATLAS}
\end{figure*}

\begin{figure*}
\begin{center}
\resizebox{0.85\textwidth}{!}{%
\includegraphics{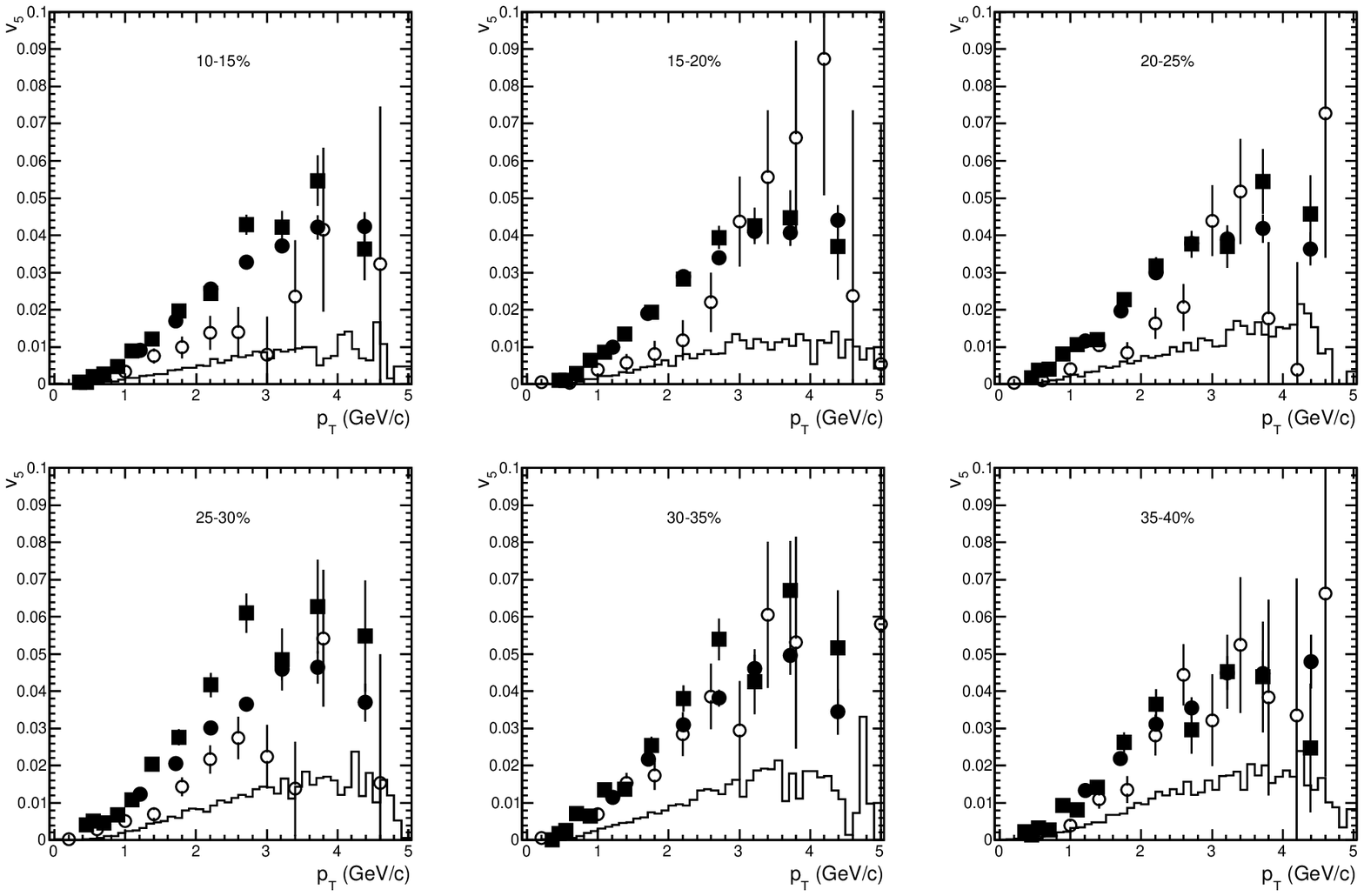}
}
\end{center}
\caption{Pentagonal flow $v_5(p_{\rm T})$ of charged hadrons at pseudo-rapidity $|\eta|<0.8$ for 
different centralities of PbPb collisions at $\sqrt s_{\rm NN}=2.76$ TeV. The closed points are 
CMS data~\cite{Chatrchyan:2013kba} ($v_5\{2\}$ --- circles, $v_5\{\rm EP\}$ --- squares), 
open circles and histograms represent $v_5\{\rm EP\}$ and $v_5(\Psi_3^{\rm RP})$ for HYDJET++ 
events, respectively.} 
\label{v5_CMS}
\end{figure*}

\begin{figure*}
\begin{center}
\resizebox{0.85\textwidth}{!}{%
\includegraphics{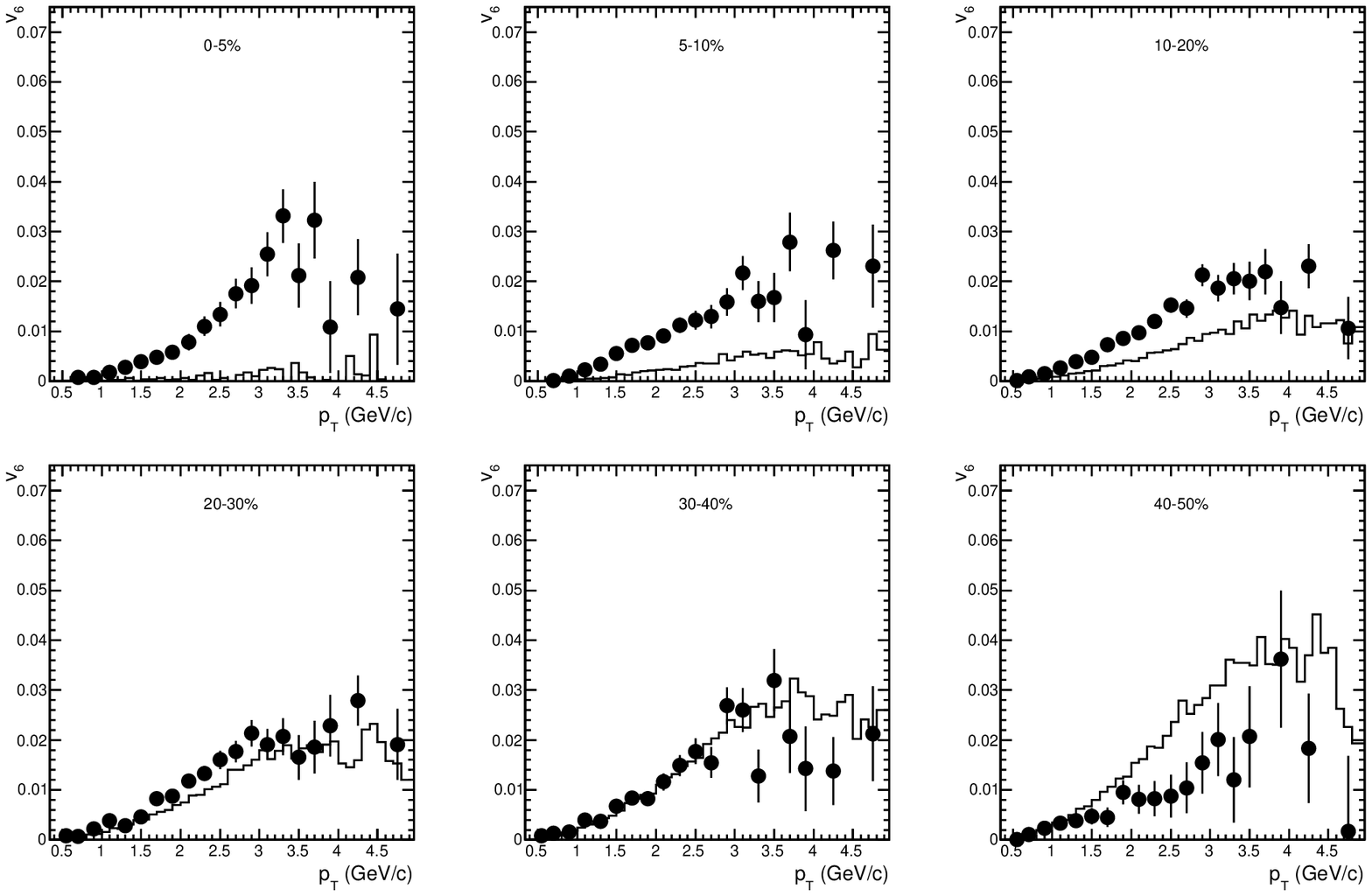}
}
\end{center}
 \caption{Hexagonal flow $v_6(p_{\rm T})$ of charged hadrons at pseudo-rapidity $|\eta|<2.5$ for 
different centralities of PbPb collisions at $\sqrt s_{\rm NN}=2.76$ TeV. The closed circles 
are ATLAS data~\cite{Aad:2012bu} on $v_6\{\rm EP\}$, histograms represent 
$v_6(\Psi_2^{\rm RP})$ for HYDJET++ events.} 
\label{v6_ATLAS}
\end{figure*}

\begin{figure*}
\begin{center}
\resizebox{0.85\textwidth}{!}{%
\includegraphics{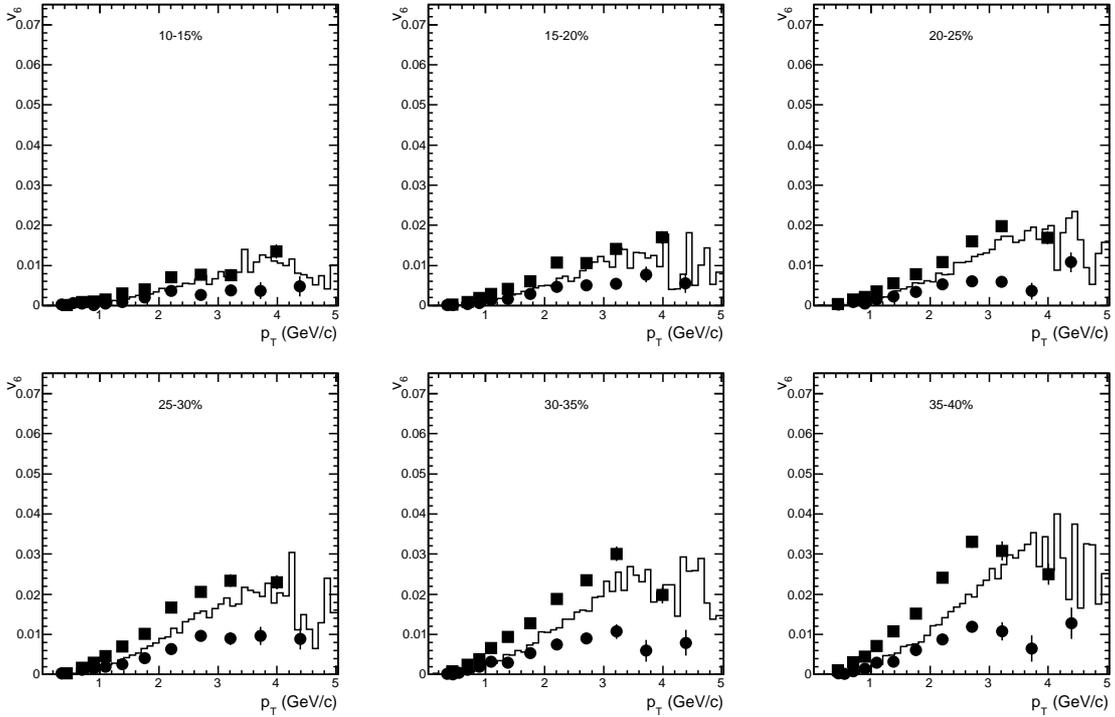}
}
\end{center}
\caption{Hexagonal flow $v_6(p_{\rm T})$ of charged hadrons at pseudo-rapidity $|\eta|<0.8$ for 
different centralities of PbPb collisions at $\sqrt s_{\rm NN}=2.76$ TeV. The closed points are 
CMS data~\cite{Chatrchyan:2013kba} ($v_6\{\rm EP/\Psi_2\}$ --- circles, $v_6\{\rm LYZ\}$ --- 
squares), histograms represent $v_6(\Psi_2^{\rm RP})$ for HYDJET++ events.} 
\label{v6_CMS}
\end{figure*}

Figures~\ref{v2_ATLAS}-\ref{v6_CMS} show anisotropic flow coefficients 
$v_{\rm n}$ as a function of the hadron transverse momentum $p_{\rm T}$. 
Let us discuss first the results of HYDJET++ simulations. It can be 
separated in two groups: (i) results obtained with respect to the true 
reaction plane straight from the generator, i.e., $v_{2,4,6}(\Psi_2^{\rm RP})$ 
and $v_{3,5}(\Psi_3^{\rm RP})$, and (ii) those obtained by using the 
(sub)event plane method with rapidity gap $|\Delta \eta|>3$. The last method 
provides us with $v_{\rm n}\{\rm EP\}$. The main systematic uncertainties for 
the methods come from non-flow correlations and flow fluctuations. The last one 
(as it is kept in the model currently) almost does not affect mean $v_{\rm n}$ 
values restored by the EP method, while the non-flow correlations can be 
effectively suppressed by applying $\eta$-gap in $v_{\rm n}$ reconstruction. 
This gives us a good reconstruction precision for elliptic $v_2$, triangular $v_3$, 
and quadrangular $v_4$ flows up to $p_{\rm T} \sim 5$ GeV/$c$. At 
higher transverse momenta some differences appear due to non-flow effects from 
jets. However, Figs.~\ref{v5_ATLAS} and \ref{v5_CMS} show that pentagonal 
flow $v_5$ determined from the model w.r.t. $\Psi_3^{\rm RP}$ and $v_5$ 
restored w.r.t. the event plane of 5-th order $\Psi_5^{\rm EP}$ differ a 
lot. The reason is that although no intrinsic $\Psi^{\rm RP}_5$ is generated
in HYDJET++, pentagonal flow $v_5$ emerges here as a result of the 
``interference'' between $v_2$ and $v_3$, each is determined with respect 
to its own reaction plane, $v_5 \propto v_2 (\Psi_2^{\rm RP}) \cdot 
v_3 (\Psi_3^{\rm RP})$, in line with the conclusions of Ref.~\cite{Teaney:2012ke}. 
Hexagonal flow $v_6$ is also very sensitive to the methods used due to nonlinear interplay 
of elliptic and triangular flows generating $v_6$, see~\cite{Bravina:2013ora} for details. 
The results of HYDJET++ for $v_6\{\rm EP\}$ are not shown on the plots because of too large 
statisitcal errors.

Note, that the experimental situation is even more complicated, and the dependence
of measured $v_{\rm n}$ on methods applied may be more crucial for all $n$ due to 
apparently larger fluctuations in the data than in the model. For instance, it was 
shown in~\cite{Heinz:2013bua} that event-by-event fluctuations in the initial 
state may lead to characteristically different $p_{\rm T}$-dependencies for the 
anisotropic flow coefficients extracted by different experimental methods.

It is also worth mentioning here that the hump-like structure of the 
simulated $v_2(p_{\rm T})$ and $v_3(p_{\rm T})$ signals appears due to 
interplay of hydrodynamics and jets. At transverse momenta $p_{\rm T}
\geq 3$\,GeV/$c$ the spectrum of hadrons is dominated by jet particles
which carry very weak flow. Thus, the elliptic and triangular flows in
the model also drop at certain $p_{\rm T}$. Higher flow harmonics arise 
in the model solely due to the presence of the $v_2$, $v_3$ and its
interference. Therefore, transverse momentum distributions of these 
harmonics inherit the characteristic hump-like shapes.

Now let us consider the ATLAS~\cite{Aad:2012bu} and 
CMS~\cite{Chatrchyan:2012ta,Chatrchyan:2013kba} data plotted onto the model results in 
Figs.~\ref{v2_ATLAS}-\ref{v6_CMS} for different centrality classes. The event plane 
for $v_{\rm n}\{\rm EP\}$ was defined experimentally with respect to n-th harmonics in all 
cases with the exeption of CMS data for $v_6\{\rm EP/\Psi_2\}$, which was measured  
using second harmonics. One can see that HYDJET++ 
reproduces experimentally measured $p_{\rm T}$-depen\-den\-ces of $v_2$, $v_3$ and 
$v_4\{{\rm LYZ}\}$ up to $p_{\rm T} \sim 5$ GeV/$c$. The centrality dependence of $v_4$ 
measured by event plane and two-particle cumulant methods is significantly weaker than that 
of $v_4$ measured by Lee-Yang zero method, presumably due to large non-flow contribution and 
increase of the flow fluctuations in more central events. Since the model is tuned to fit the 
$p_{\rm T}-$dependencies of $v_{4}\{{\rm LYZ}\}$, it underestimates the quadrangular flow, 
restored by the EP or two-particle cumulant methods, in (semi-)central collisions.
Recall, that in ideal hydrodynamics (at the limit of small temperatures, large transverse 
momenta and absence of the flow fluctuations) $v_4\{\Psi_2\} / v_2^2 = 0.5$~\cite{Borghini:2005kd}.

The same trend is seen for $p_{\rm T}$-dependencies of the pentagonal flow. 
For central and semi-central topologies up to $\sigma/\sigma_{\rm geo} \approx 20\%$ 
the $v_5\{{\rm EP}\}$ in the model underestimates the experimentally measured 
$v_5\{{\rm EP}\}$, whereas for more peripheral collisions the 
agreement between the model and the data is good. Unfortunately, there 
are no data on pentagonal flow extracted by the LYZ method. As we have 
seen, for $v_2,\ v_3$ and $v_4$ in central and semi-central collisions
the LYZ method provides noticeably weaker flow compared to that obtained 
by the EP method. One may expect, therefore, that the pentagonal flow, 
$v_5\{{\rm LYZ}\}$, almost free from non-flow contributions, should be closer to the 
$v_5$ generated by the HYDJET++. If the future experimental data on $v_5$ will 
persist on stronger flow, this fact can be taken as indication of the
possible presence of additional pentagonal eccentricity $\epsilon_5(b)$ 
with the new phase $\Psi_5^{\rm RP}$ responsible for genuine $v_5$. Both
parameters can be easily inserted in Eq.~(\ref{Rbphi}) for the modulation 
of the final freeze-out hypersurface.

At last, $p_{\rm T}$-dependencies of the hexagonal flow in HYDJET++ are similar to that seen  
in CMS data within the uncertainties related to methods used. However  
$v_{6}(\Psi_2^{\rm RP})$ in the model visibly underestimates ATLAS data on $v_6\{\rm EP\}$
for most central events. The latter fact may be explained by a siginificant 
$v_3$ contribution to $v_6\{\rm EP\}$ in central collisions, which is not presented in 
$v_{6}(\Psi_2^{\rm RP})$ component: 
$v_{6}(\Psi_3^{\rm RP}) \sim v_{6}(\Psi_2^{\rm RP}) < v_6\{\rm EP\}$.
On the other hand, the relative contribution to $v_6\{\rm EP\}$ coming from $v_2$ is 
instantly increasing as the reaction becomes more peripheral~\cite{Bravina:2013ora}, and 
starting from $20-30$\% centralities we already get $v_6\{{\rm EP}\} \sim v_{6}(\Psi_2^{\rm RP}) 
\gg v_{6}(\Psi_3^{\rm RP})$ with the approximate agreement between the model and the data. 

Some additional checks have been done as well. In the presence of only
elliptic flow all odd higher harmonics are found to be essentially zero.
The quadrangular flow is zero, $v_4 = 0$, if the elliptic flow is absent.
The pentagonal flow disappears, $v_5 = 0$, in case of either $v_2 = 0$ 
or $v_3 = 0$. The hexagonal flow is zero, $v_6 = 0$, if both 
elliptic and triangular flows are absent, $v_2 = 0$ and $v_3 = 0$.

\subsection{Anisotropy harmonics for identified charge hadrons}
\label{subsec:res2}

\begin{figure*}
\begin{center}
\resizebox{0.85\textwidth}{!}{%
\includegraphics{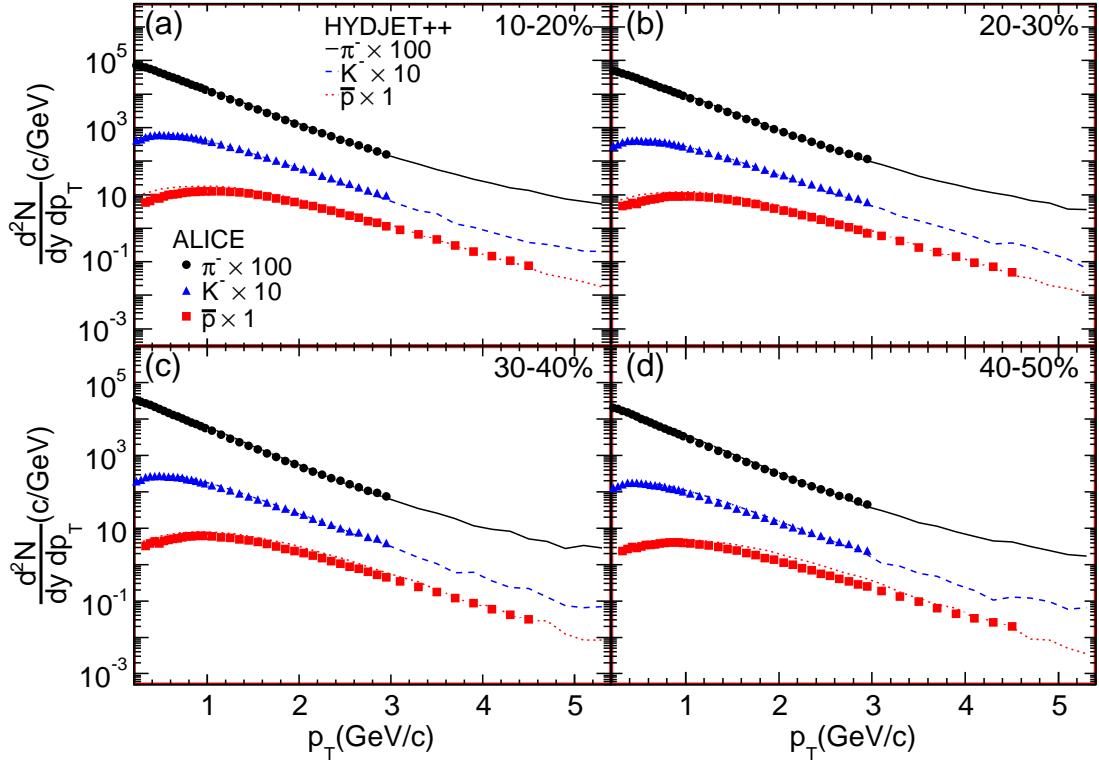}
}
\end{center}
 \caption{Negatively charged pion, kaon and anti-proton transverse momentum spectra at 
pseudo-rapidity $|\eta|<0.5$ for different centralities of PbPb collisions at 
$\sqrt s_{\rm NN}=2.76$ TeV. The points are ALICE data~\cite{Preghenella:2011np}, 
histograms are the simulated HYDJET++ events. The spectra
are scaled by different factors for visual convenience.} 
\label{dndpt-pid}
\end{figure*}

Finally, let us consider distributions for some hadronic species measured in PbPb collisions 
at the LHC. Before addressing to azimuthal anisotropy harmonics of identified hadrons, the 
comparision of HYDJET++ results with ALICE data~\cite{Preghenella:2011np} on 
$p_{\rm T}$-spectra of negatively charged pions, kaons and anti-protons in PbPb collisions 
is displayed in Fig.~\ref{dndpt-pid}. One can see that HYDJET++ reproduces well the measured 
transverse momentum spectra of identified hadrons within the whole range of accessible 
$p_{\rm T}$.  

\begin{figure*}
\begin{center}
\resizebox{1.05\textwidth}{!}{%
\includegraphics{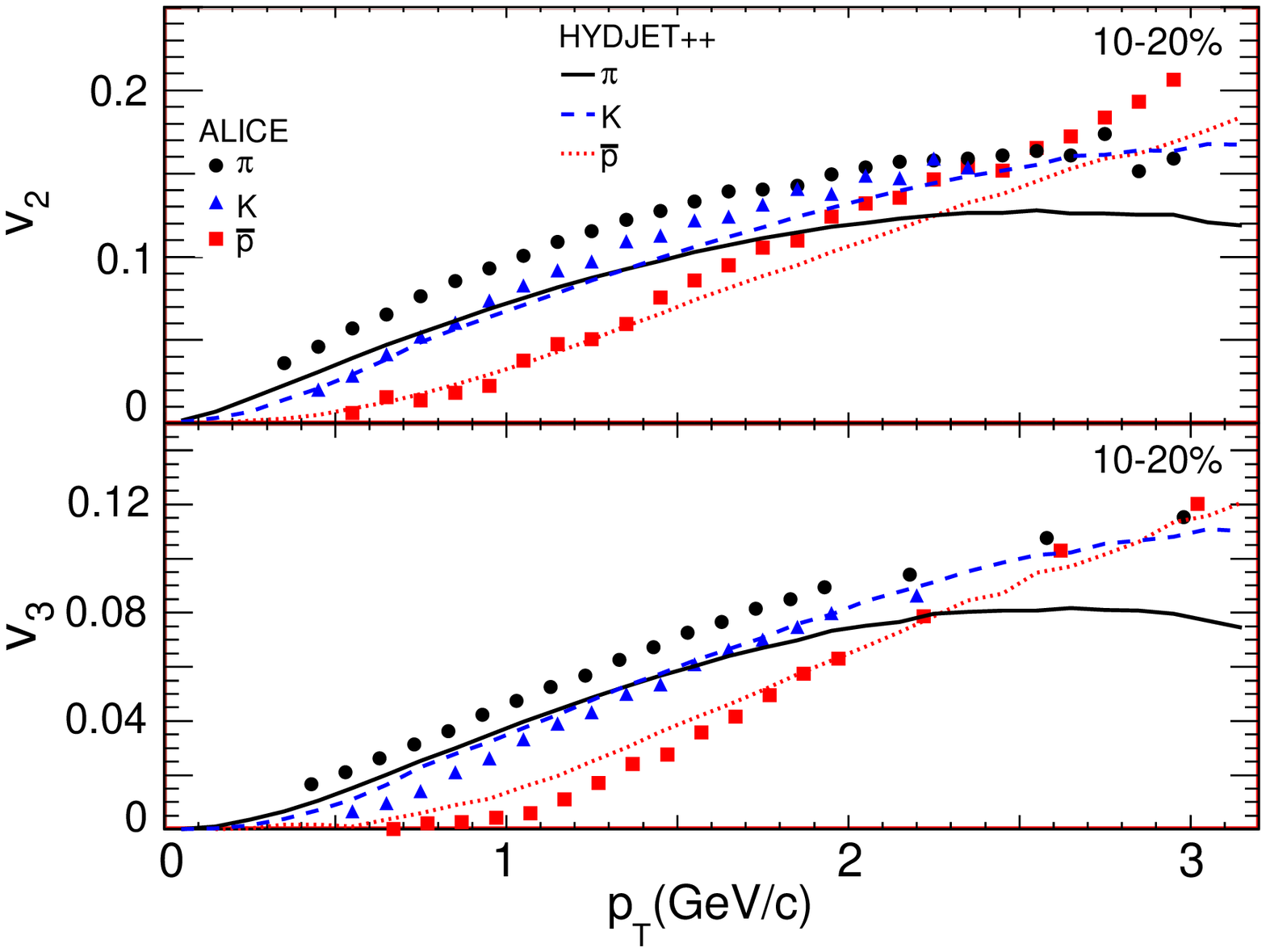} 
\includegraphics{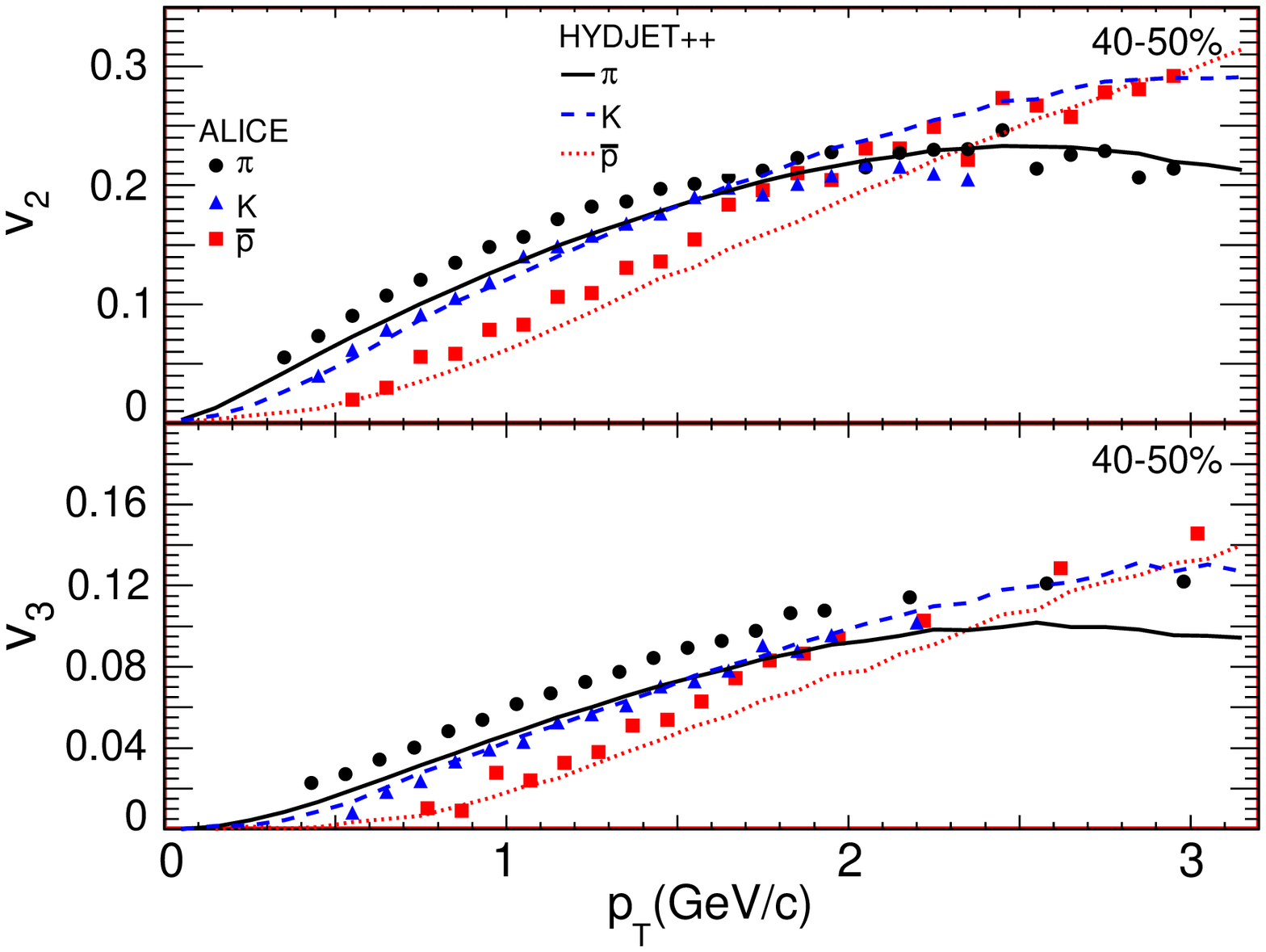}
}
\end{center}
 \caption{Elliptic $v_2(p_{\rm T})$ and triangular $v_3(p_{\rm T})$ flows of charged pions, kaons and 
anti-protons at pseudo-rapidity $|\eta|<0.8$ for 10--20\% (left) and 40--50\% (right) 
centrality of PbPb collisions at $\sqrt s_{\rm NN}=2.76$ TeV. The points are ALICE 
data~\cite{Krzewicki:2011ee}, histograms are the simulated HYDJET++ events.} 
\label{v2v3-pid}
\end{figure*}

Figure~\ref{v2v3-pid} presents the comparision of HYDJET++ results and the ALICE 
data~\cite{Krzewicki:2011ee} for the elliptic and triangular flow of pions, kaons and 
anti-protons at 10--20\% and 40--50\%  centrality of PbPb collisions. The agreement between 
the model and the data for kaons and anti-protons looks fair. For pions the
model underestimates the data a bit. The discrepancy is more pronounced for more central 
collisions indicating, perhaps, presence of strong non-flow correlations in the data.

\section{Conclusion}
\label{sec:summ}

Azimuthal anisotropy harmonics of inclusive and identified charged hadrons in PbPb collisions 
at $\sqrt{s}_{\rm NN}=2.76$ TeV have been analyzed in the framework of HYDJET++ model. 
The effects of possible non-elliptic shape of the initial overlap of the
colliding nuclei are implemented in HYDJET++ by the modulation of the final freeze-out
hypersurface with the appropriate fitting triangular coefficient. This modulation is not 
correlated with the direction of the impact parameter, and two independent ``strong'' 
lower azimuthal harmonics, $v_2$ and $v_3$, being obtained as a result. They are of 
different physical origin, coded partly in the different centrality dependence. 
Interference between $v_2$ and $v_3$ generates as ``overtones'' both even and odd higher 
azimuthal harmonics, $v_4$, $v_5$, $v_6$, etc.

This mechanism allows HYDJET++ to reproduce the LHC data on $p_{\rm T}$- and centrality 
dependencies of the aniso\-t\-ro\-pic flow coefficients $v_n$ (n=2$\div$4) up to $p_{\rm T} 
\sim 5$ GeV/$c$ and $40$\% centrality, and also the basic trends for pentagonal $v_5$ and 
hexagonal $v_6$ flows. Some discrepancy between the model results and the data on the 
pentagonal flow in central events requires further study of additional sources of the 
non-flow correlations and flow fluctuations, which may be absent in the model. Although the introduction of internal higher harmonics is also possible in the HYDJET++, there is no clear
evidence in the data to do so at present. Obtained results show that higher harmonics of the
azimuthal flow get very significant contributions from the lower harmonics, $v_2$ and $v_3$. 
This circumstance makes it difficult to consider the higher harmonics as independent
characteristics of the early phase of ultrarelativistic heavy ion collisions.

\section*{Acknowledgments}

\begin{acknowledgement}
Discussions with A.V.~Belyaev and D.~d'Enterria are gratefully acknowledged.
We thank our colleagues from CMS, ALICE and ATLAS collaborations for 
fruitful cooperation. This work was supported by Russian Foundation for 
Basic Research (grant 12-02-91505), Grants of President of Russian Federation 
for Scientific Schools Supporting (No. 3920.2012.2 and No. 3042.2014.2), Ministry 
of Education and Sciences of Russian Federation (agreement No. 8412), Norwegian Research
Council (NFR), European Union and the Government of Czech Republic under 
the project "Support for research teams on CTU" (No. CZ.1.07/2.3.00/30.0034).
\end{acknowledgement}

\begin{thebibliography}{99}

\bibitem{d'Enterria:2006su} D.~d'Enterria, J. Phys. {\bf G 34}, (2007) S53
\bibitem{Hwa:2010} {\it Quark Gluon Plasma 4}, edited by R.C. Hwa and 
X.-N. Wang (World Scientific, Singapore, 2010).
\bibitem{Salgado:2009jp} C.~Salgado, Proceedings of European School of 
High-Energy Physics (2008) 239, arXiv:0907.1219 [hep-ph]
\bibitem{Dremin:2010jx} I.M.~Dremin, A.V.~Leonidov, Phys. Usp. {\bf 53}, (2011) 1123
\bibitem{Muller:2012zq} B.~Muller, J.~Schukraft, B.~Wyslouch, Ann. Rev. Nucl. Part. Sci. 
{\bf 62}, (2012) 361
\bibitem{Aamodt:2010pa} K.~Aamodt, et al. (ALICE Collaboration), Phys. Rev. Lett. {\bf 105}, 
(2010) 252302
\bibitem{ATLAS:2011yk} G.~Aad, et al. (ATLAS Collaboration), Phys. Lett. {\bf B 707}, (2012) 330
\bibitem{Chatrchyan:2012ta} S.~Chatrchyan, et al. (CMS Collaboration),  Phys. Rev. {\bf C 87}, 
(2013) 014902
\bibitem{ALICE:2011ab} K.~Aamodt, et al. (ALICE Collaboration), Phys. Rev. Lett. {\bf 107}, 
(2011) 032301
\bibitem{Aad:2012bu} G.~Aad, et al. (ATLAS Collaboration), Phys. Rev. {\bf C 86}, (2012) 014907
\bibitem{Chatrchyan:2013kba} S.~Chatrchyan, et al. (CMS Collaboration), arXiv:1310.8651 [nucl-ex]
\bibitem{Abelev2012:di} B.~Abelev, et al. (ALICE Collaboration), Phys. Lett. {\bf B 719}, 
(2013) 18
\bibitem{Kolb:2003zi} P.F.~Kolb, Phys. Rev. {\bf C 68}, (2003) 031902 
\bibitem{Kolb:2004gi} P.F.~Kolb, L.-W.~Chen, V.~Greco, C.M.~Ko, Phys. Rev. {\bf C 69}, (2004)
 051901 
\bibitem{Alver:2010gr} B.~Alver, G.~Roland, Phys. Rev. {\bf C 81}, (2010) 054905 
\bibitem{Alver:2010dn} B.H.~Alver, C.~Gombeaud, M.~Luzum, J.Y.~Ollitrault, Phys. Rev. 
{\bf C 82}, (2010) 034913 
\bibitem{Ollitrault:1992bk} J.Y.~Ollitrault, Phys. Rev. {\bf D 46}, (1992) 229
\bibitem{Gyulassy:2000gk} M.~Gyulassy, I.~Vitev, X.-N. Wang, Phys. Rev. Lett. {\bf 86}, (2001) 
2537
\bibitem{Lokhtin:2012re} I.P.~Lokhtin, A.V.~Belyaev, L.V~Malinina, S.V.~Petrushanko, 
E.P.~Rogochaya, A.M.~Snigirev, Eur. Phys. J. {\bf C 72}, (2012) 2045 
\bibitem{Lokhtin:2008xi} I.P.~Lokhtin, L.V~Malinina, S.V.~Petrushanko, A.M.~Snigirev,
I.~Arsene, K.~Tywoniuk, Comput. Phys. Commun. {\bf 180}, (2009) 779
\bibitem{Bravina:2013upa} L. Bravina, B.H. Brusheim Johansson, G. Eyyubova, E. Zabrodin,
Phys. Rev. {\bf C 87}, (2013) 034901 
\bibitem{Noferini:2012ps} F.~Noferini (for the ALICE Collaboration), Nucl. Phys.  
{\bf A 904-905}, (2013) 438c
\bibitem{Eyyubova:2009hh} G.~Eyyubova, et al., Phys. Rev. {\bf C 80}, (2009) 064907
\bibitem{Bravina:2013ora} L.~Bravina, et al., arXiv:1311.0747 [hep-ph]
\bibitem{Xu:2011jm} J.~Xu, C.M.~Ko, Phys. Rev. {\bf C 84}, (2011) 044907
\bibitem{Gale:2012rq} C.~Gale, S.~Jeon, B.~Schenke, P.~Tribedy, R.~Venugopalan, 
Phys. Rev. Lett. {\bf 110}, (2013) 012302
\bibitem{Lokhtin:2005px} I.P.~Lokhtin, A.M.~Snigirev, Eur. Phys. J. {\bf C 45}, (2006) 211
\bibitem{Amelin:2006qe} N.S.~Amelin et al., Phys. Rev. {\bf C 74}, (2006) 064901
\bibitem{Amelin:2007ic} N.S.~Amelin et al., Phys. Rev. {\bf C 77}, (2008) 014903
\bibitem{Wiedemann:1997cr} U.~Wiedemann, Phys. Rev. {\bf C 57}, (1998) 266
\bibitem{Sjostrand:2006za} T.~Sjostrand, S.~Mrenna, P.~Skands, JHEP {\bf 0605}, (2006) 026
\bibitem{Tywoniuk:2007xy} K.~Tywoniuk, I.C.~Arsene, L.~Bravina, A.B.~Kaidalov,
E.~Zabrodin, Phys. Lett. {\bf B 657}, (2007) 170
\bibitem{Poskanzer:1998yz} A.M.~Poskanzer, S.A.~Voloshin, Phys. Rev. {\bf C 58}, (1998) 1671
\bibitem{Borghini:2001vi} N.~Borghini, P.M.~Dinh, J.Y.~Ollitrault, Phys. Rev. {\bf C 64}, 
(2001) 054901
\bibitem{Borghini:2001zr} N.~Borghini, P.M.~Dinh, J.Y.~Ollitrault, nucl-ex/0110016 
\bibitem{Bhalerao:2003xf} R.S.~Bhalerao, N.~Borghini, J.Y.~Ollitrault, Nucl. Phys. {\bf A 727}, 
(2003) 373 
\bibitem{Borghini:2004ke} N.~Borghini, R.S.~Bhalerao, J.Y.~Ollitrault, J. Phys. {\bf G 30}, 
(2004) S1213
\bibitem{Teaney:2012ke} D.~Teaney, L.~Yan, Phys. Rev. {\bf C 86}, (2012) 044908
\bibitem{Heinz:2013bua} U.~Heinz, Z.~Qiu, C.~Shen, Phys. Rev. {\bf C 87}, (2013) 034913
\bibitem{Borghini:2005kd} N.~Borghini, J.-Y.~Ollitrault, Phys. Lett. {\bf B 642}, (2006) 227
\bibitem{Preghenella:2011np} R.~Preghenella (for the ALICE Collaboration), Acta Phys. Polon. 
{\bf B 43}, (2012) 555
\bibitem{Krzewicki:2011ee} M.~Krzewicki (for the ALICE Collaboration), J. Phys. {\bf G 38}, 
(2011) 124047
\end{thebibliography}
\end{document}